\documentclass[12pt]{article}

\newcommand{\br}{{\bf R}}

\newcommand{\Om}{\Omega}

\newcommand{\Tr}{\rm{Tr}}

\usepackage{latexsym,amsfonts,amsmath,amsthm,amssymb,graphicx}

\usepackage[T1]{fontenc}
\usepackage[latin1]{inputenc}

\date{}
\title{INTERPRETATION OF STATIONARY STATES IN PREQUANTUM CLASSICAL
STATISTICAL FIELD THEORY}
\author{Andrei Khrennikov\\
International Center for
Mathematical Modeling \\
in Physics and Cognitive Sciences\\
University of V\"axj\"o, Sweden\\
Email:Andrei.Khrennikov@msi.vxu.se}

\begin{document}
\maketitle

 \abstract{We develop a prequantum classical statistical
model in that the role of hidden variables is played by classical
(vector) fields. We call this model Prequantum Classical Statistical
Field Theory (PCSFT). The correspondence between classical and
quantum quantities is asymptotic, so we call our approach asymptotic
dequantization. In this note we pay the main attention to
interpretation of so called pure quantum states (wave functions) in
PCSFT, especially stationary states. We show, see Theorem 2, that
pure states of QM can be considered as  labels for Gaussian measures
concentrated on one dimensional complex subspaces of phase space
that are invariant with respect to the Schr\"odinger dynamics. ``A
quantum system in a stationary state $\psi$'' in PCSFT is nothing
else than a Gaussian ensemble of classical fields (fluctuations of
the vacuum field of a very small magnitude) which is not changed in
the process of Schr\"odinger's evolution. We interpret in this way
the problem of {\it stability of hydrogen atom.}

\bigskip

\bigskip


\section{INTRODUCTION}

The problem of {\it completeness of QM} has been an important source
of investigations on quantum foundations, see, e.g., for recent
debates Ref. [1]-[6]. Now days this problem is typically regarded as
the problem of {\it hidden variables.} This problem is not of purely
philosophic interest. By constructing a model that would provide a
finer description of physical reality than given by the quantum wave
function $\psi$ we obtain at least theoretical possibility to go
{\it beyond quantum mechanics.} In principle, we might find effects
that are not described by quantum mechanics. One of the main
barriers on the way beyond quantum mechanics are various ``NO-GO''
theorems (e.g., theorems of von Neumann, Kochen-Specker, Bell,...).
Therefore by looking for a prequantum classical statistical model
one should take into account all known ``NO-GO'' theorems.

In a series of papers [7] there was shown that in principle all
distinguishing features of quantum probabilities (e.g., {\it
interference, Born's rule}, representation of random variables by
noncommuting operators) can be obtained in classical (but
contextual) probabilistic framework. The main problem was to find a
classical statistical model which would be natural from the physical
viewpoint. One of such models was presented in [8]. It was shown
that it is possible to represent quantum mechanics as an asymptotic
projection of classical statistical mechanics on {\it
infinite-dimensional phase space} $\Omega= H\times H,$ where $H$ is
Hilbert space. By realizing Hilbert space $H$ as the $L_2({\bf
R}^3)$-space we obtain the representation of prequantum classical
phase space as the space of classical (real vector) fields
$\psi(x)=(q(x), p(x))$ on ${\bf R}^3.$ We call this approach to the
problem of hidden variables {\it Prequantum Classical Statistical
Field Theory,} PCSFT. In this model quantum states are just labels
for Gaussian ensembles of classical fields. Such ensembles (Gaussian
measures $\rho)$ are characterized by zero mean value and very small
dispersion:
\begin{equation}
\label{SD} \int_{L_2({\bf R}^3)
 \times L_2({\bf R}^3)} \int_{{\bf
R}^3} [p^2(x) + q^2(x)] dx d\rho (q,p) = \alpha, \; \alpha\to 0.
\end{equation}
This dispersion is a small parameter of the model. Quantum mechanics
is obtained as the $\lim_{\alpha\to 0}$ of PCSFT.

Let us consider the ``classical vacuum field.'' In PCSFT it is
represented by the function $\psi_{\rm{vacuum}}\equiv 0.$ Since a
Gaussian ensemble of classical fields has the zero mean value, these
fields can be considered as random fluctuations of the ``classical
vacuum field.'' Since dispersion is very small, these are very small
fluctuations. There is some similarity with SED and stochastic QM,
cf. [9]. The main difference is that we consider fluctuations not on
``physical space'' ${\bf R}^3,$ but on infinite dimensional space of
classical fields.

In [8] we studied asymptotic expansions of Gaussian integrals of
analytic functionals and obtained an asymptotic equality coupling
the Gaussian integral  and the trace of the composition of scaling
of the covariation operator of a Gaussian measure and the second
derivative of a functional. In this way we coupled the classical
average (given by an infinite-dimensional Gaussian integral) and the
quantum average (given by the von Neumann trace formula). In [8]
there was obtained generalizations of QM that were based on
expansions of classical field-functionals into Taylor series up to
terms of the degree $n=2, 4, 6,..$ (for $n=2$ we obtain the ordinary
QM).

In the present paper we change crucially the interpretation of the
small parameter of our model. In [8] this parameter was identified
with the Planck constant $h$ (in making such a choice I was very
much stimulated by discussions with people working in SED and
stochastic quantum mechanics, cf. [9]).  In this we paper consider
$\alpha$ as a new parameter  giving the dispersion of prequantum
fluctuations. We construct a one parameter family of classical
statistical models $M^\alpha,\; \alpha \geq 0.$ QM is obtained as
the limit of classical statistical models when $\alpha \to 0:$
\begin{equation} \label{CP}
 \lim_{\alpha\to 0}M^\alpha=N_{\rm{quant}},
\end{equation}
where $N_{\rm{quant}}$ is the Dirac-von Neumann quantum model [10],
[11]. Our approach should not be mixed with {\it deformation
quantization,} see, e.g., [12]. In the formalism of deformation
quantization classical mechanics on the phase-space $\Omega_{2n} =
{\bf R}^{2n}$ is obtained as the $\lim_{h\to 0}$ of quantum
mechanics (the correspondence principle). In the deformation
quantization the quantum model is considered as depending on a small
parameter $h:N_{\rm{quant}}\equiv N_{\rm{quant}}^h,$ and formally
\begin{equation}
\label{CP1}
 \lim_{h \to 0} N_{\rm{quant}}^h = M_{\rm{conv. class.}}
\end{equation}
where  $M_{\rm{conv. class.}}$ is the conventional classical model
with the phase-space $\Omega_{2n}.$

The main problem is that our model does not provide the magnitude of
$\alpha.$ We may just speculate that there might be some relations
with scales of quantum gravity and string theory.

In this article we pay the main attention to the interpretation of
so called {\it pure states} in PCSFT, especially so called {\it
stationary states.} We show, see Theorem 2, that pure states of QM
can be interpreted as simply labels for Gaussian measures
concentrated on one dimensional complex subspaces of phase space
that are invariant with respect to the Schr\"odinger dynamics. Thus
PCSFT implies the following viewpoint to quantum stationarity. First
of all this is not deterministic classical stationarity.
Nevertheless, this is purely classical, but stochastic stationarity,
cf. [13]. ``A quantum system in a stationary state $\psi$'' in PCSFT
is nothing else than a Gaussian ensemble of classical fields
(fluctuations of the vacuum field of a very small magnitude) which
is not changed in the process of Schr\"odinger's evolution. We
interpret in this way the problem of {\it stability of hydrogen
atom,} see section 7. Here ``an electron on a stationary orbit'' is
a stationary Gaussian ensemble of classical fields. The structure of
these Gaussian fluctuations provides the picture of a {\it bound
state.}

To simplify the introduction to PCSFT, in papers [8] we considered
quantum models over the real Hilbert space and only in section 5 of
the second paper in [8] there were given main lines of
generalization to the complex Hilbert space. In this paper we start
directly with the complex case. Here the crucial role is played by
the symplectic structure on the infinite-dimensional phase space
$\Omega,$ cf. [12]. In particular, in our model all classical
physical variables should be invariant with respect to the
symplectic operator $J,\; J^2=-I.$

We show that the Schr\"odinger dynamics is nothing else than
Hamilton dynamics on $\Omega.$ Therefore quantum stationary states
can be considered as invariant measures (concentrated on
$J$-invariant planes of phase space $\Omega)$ of special
infinite-dimensional Hamiltonian systems.

In contrast to [8], in this paper we study asymptotic of classical
averages (given by Gaussian functional integrals) on the matheatical
level of rigor. We find a correct functional class in that such
expansions are valid and obtain an estimate of the rest term in the
fundamental asymptotic formula coupling classical and quantum
averages.

\section{ASYMPTOTIC DEQUANTIZATION}

We define {\it ``classical statistical models''} in the following
way, see [8] for more detail (and even philosophic considerations):
a) physical states $\omega$ are represented by points of some set
$\Omega$ (state space); b) physical variables are represented by
functions $f: \Omega \to {\bf R}$ belonging to some functional space
$V(\Omega);$ c) statistical states are represented by probability
measures on $\Omega$ belonging to some class $S(\Omega);$ d) the
average of a physical variable (which is represented by a function
$f \in V(\Omega))$ with respect to a statistical state (which is
represented by a probability measure  $\rho \in S(\Omega))$ is given
by
\begin{equation}
\label{AV0} < f >_\rho \equiv \int_\Omega f(\psi) d \rho(\psi) .
\end{equation}

A {\it classical statistical model} is a pair $M=(S, V).$ We recall
that classical statistical mechanics on the phase space
$\Omega_{2n}= {\bf R}^n\times {\bf R}^n$ gives an example of a
classical statistical model. But we shall not be interested in this
example in our further considerations. We shall develop  a classical
statistical model with {\it an infinite-dimensional phase-space.}

The conventional quantum statistical model with the complex Hilbert
state space $\Omega_c$ is described in the following way (see
Dirac-von Neumann [10], [11] for the conventional complex model): a)
physical observables are represented by operators $A: \Omega_c \to
\Omega_c$ belonging to the class of continuous self-adjoint
operators ${\cal L}_s \equiv {\cal L}_s (\Omega_c);$ b) statistical
states are represented by von Neumann density operators, see [4]
(the class of such operators is denoted by ${\cal D} \equiv {\cal D}
(\Omega_c));$ d) the average of a physical observable (which is
represented by the operator $A \in {\cal L}_s (\Omega_c))$ with
respect to a statistical state (which is represented
  by the density operator $D \in {\cal D} (\Omega_c))$ is given by von Neumann's
formula [11]:
\begin{equation}
\label{AV1} <A >_D \equiv \rm{Tr}\; DA
\end{equation}
The {\it quantum statistical model} is the pair $N_{\rm{quant}}
=({\cal D}, {\cal L}_s).$

We are looking for a classical statistical model $M=(S, V)$ which
will give ``dequantization'' of the quantum model $N_{\rm{quant}}
=({\cal D}, {\cal L}_s).$ Here the meaning of ``dequantization''
should be specified. In fact, all ``NO-GO'' theorems (e.g., von
Neumann, Kochen-Specker, Bell,...) can be interpreted as theorems
about impossibility of various dequantization procedures. Therefore
we should define the procedure of dequantization in such a way that
there will be no contradiction with known ``NO-GO'' theorems, but
our dequantization procedure still will be natural from the physical
viewpoint. We define (asymptotic) dequantization as a family
$M^\alpha=(S^\alpha, V)$ of classical statistical models depending
on small parameter $\alpha \geq 0.$ There  should exist maps
$T:S^\alpha\to {\cal D}$ and $T: V \to {\cal L}_s$ such that: a)
both maps are {\it surjections} (so all quantum objects are covered
by classical); b) the map $T: V \to {\cal L}_s$ is ${\bf R}$-linear
(we recall that we consider real-valued classical physical
variables); c) the map $T:S\to {\cal D}$ is injection (there is
one-to one correspondence between classical and quantum statistical
states); d) classical and quantum averages are coupled through the
following asymptotic equality:
 \begin{equation}
\label{AQ} < f >_\rho = \alpha <T(f)>_{T(\rho)} + o(\alpha), \; \;
\alpha \to 0
\end{equation}
(here $<T(f)>_{T(\rho)}$ is the  quantum average); so:
\begin{equation}
\label{AQ1} \int_\Omega f(\psi) d \rho(\psi)=  \alpha \; \Tr \; D A
+ o(\alpha), \; \; A=T(f), D= T(\rho).
\end{equation}
This equality can be interpreted in the following way. Let $f(\psi)$
be a classical physical variable (describing properties of
microsystems - classical fields having very small magnitude
$\alpha).$  We define its {\it amplification} by:
 \begin{equation}
\label{AMP} f_\alpha (\psi) =\frac{1}{\alpha} f(\psi)
\end{equation}
(so any micro effect is amplified in $\frac{1}{\alpha}$-times). Then
we have:
%\begin{equation} \label{AQ4}
$< f_\alpha >_\rho = <T(f)>_{T(\rho)} + o(1), \; \; \alpha \to 0,$
%\end{equation}
or
\begin{equation} \label{AQ5}
\int_\Omega f_\alpha(\psi) d \rho(\psi)=   \Tr \; D A + o(1), \; \;
A=T(f), D= T(\rho).
\end{equation}
Thus: {\it Quantum average $\approx$ Classical average of the
$\frac{1}{\alpha}$-amplification.} Hence: {\it QM is a mathematical
formalism describing a statistical approximation of amplification of
micro effects.}

We see that for physical variables/quantum observables and classical
and quantum statistical states the dequantization  maps have
different features. The map $T: V\to  {\cal L}_s$ is not injective.
Different classical physical variables $f_1$ and $f_2$ can be mapped
into one quantum observable $A.$ This is not surprising. Such a
viewpoint on the relation between classical variables and quantum
observables was already presented by J. Bell, see [14]. In
principle, experimenter could not distinguish classical (``ontic'')
variables by his measurement devices. In contrast, the map
$T:S^\alpha\to {\cal D}$ is injection. Here we suppose that quantum
statistical states represent uniquely  (``ontic'') classical
statistical states.

The crucial difference with dequantizations considered in known
``NO-GO'' theorems is that in our case classical and quantum
averages are equal only asymptotically and that a classical variable
$f$ and the corresponding quantum observable $A=T(f)$ can have
different ranges of values.

\section{PREQUANTUM CLASSICAL STATISTICAL MODEL}

We choose the phase space $\Om= Q\times P,$ where $Q=P=H$ and $H$ is
the infinite-dimensional real (separable) Hilbert space. We consider
$\Omega$ as the real Hilbert space with the scalar product $(\psi_1,
\psi_2)= (q_1, q_2) + (p_1, p_2).$ We denote  by $J$ the symplectic
operator on $\Omega:
 J= \left( \begin{array}{ll}
 0&1\\
 -1&0
 \end{array}
 \right ).$
Let us consider the class ${\cal L}_{\rm symp} (\Omega)$ of bounded
${\bf R}$-linear operators $A: \Omega \to \Omega$ which commute with
the symplectic operator:
\begin{equation}
\label{SS} A J= J A
\end{equation}
This is a subalgebra of the algebra of bounded linear operators
${\cal L} (\Omega).$ We also consider the space of ${\cal
L}_{\rm{symp}, s}(\Omega)$ consisting of self-adjoint operators.

By using the operator $J$ we can introduce on the phase space
$\Omega$ the complex structure. Here $J$ is realized as $-i.$ We
denote $\Omega$ endowed with this complex structure by $\Omega_c:
\Omega_c\equiv Q\oplus i P.$ We shall use it later. At the moment
consider $\Omega$ as a real linear space and consider its
complexification $\Omega^{{\bf C}}= \Omega \oplus i \Omega.$

Let us consider the functional space ${\cal V}_{\rm{symp}}(\Omega)$
consisting of functions $f:\Omega \to {\bf R}$ such that:

a) the state of vacuum is  preserved : $f(0)=0;$

b) $f$ is $J$-invariant: $f(J\psi)= f(\psi);$

c) $f$ can be extended to the  analytic function $f:\Omega^{{\bf
C}}\to {\bf C}$ having  the exponential growth:
$$
\vert f(\psi)\vert \leq c_f e^{r_f \Vert \psi \Vert}
$$
for some $c_f, r_f \geq 0$ and for all $\psi\in \Omega^{{\bf C}}.$
We remark that the possibility to extend a function $f$ analytically
onto $\Omega^{{\bf C}}$ and the exponential estimate on
$\Omega^{{\bf C}}$ plays the important role in the asymptotic
expansion of Gaussian integrals. To get a mathematically rigor
formulation, conditions in [8] should be reformulated in the similar
way.

The following trivial mathematical result plays the fundamental role
in establishing classical $\to$ quantum correspondence: {\it Let $f$
be a smooth $J$-invariant function. Then } $f^{\prime \prime}(0)\in
{\cal L}_{\rm{symp}, s}(\Omega).$ In particular, a quadratic form is
$J$-invariant iff it is determined by an operator belonging to
${\cal L}_{\rm{symp}, s}(\Omega).$

We consider the space statistical states $S_{G,
\rm{symp}}^{\alpha}(\Omega)$ consisting of measures $\rho$ on
$\Omega$ such that: a) $\rho$ has zero mean value; b) it is a
Gaussian measure; c) it is $J$-invariant; d) its dispersion has the
magnitude $\alpha.$ Thus these are $J$-invariant Gaussian measures
such that $$ \int_\Omega \psi d\rho(\psi)=0 \; \mbox{and}\;
\sigma^2(\rho)= \int_\Omega \Vert \psi\Vert^2 d \rho(\psi)= \alpha,
\; \alpha \to 0.
$$
Such measures describe small Gaussian fluctuations of the vacuum
field.

The following trivial mathematical result plays the fundamental role
in establishing classical $\to$ quantum correspondence: {\it Let a
measure $\rho$ be $J$-invariant. Then its covariation operator} $B=
\rm{cov}\; \rho \in {\cal L}_{\rm{symp}, s}(\Omega).$ Here $(By_1,
y_2)= \int (y_1, \psi)(y_2, \psi) d \rho( \psi).$

We now consider the complex realization $\Omega_c$ of the phase
space and the corresponding complex scalar product $<\cdot, \cdot>.$
We remark that the class of operators ${\cal L}_{\rm symp} (\Omega)$
is mapped onto the class of ${\bf C}$-linear operators ${\cal
L}(\Omega_c).$ We also remark that, for any $A\in {\cal
L}_{\rm{symp}, s}(\Omega),$ real and complex quadratic forms
coincide:
\begin{equation}
\label{CI1}
 (A\psi,\psi) =<A\psi,\psi>.
\end{equation}
We also define for any measure its complex covariation operator
$B^c= \rm{cov}^c \rho$ by
$$
<B^c y_1, y_2>=\int <y_1, \psi> <\psi, y_2> d \rho (\psi).
$$
We remark that for a $J$-invariant measure $\rho$ its complex and
real covariation operators are related as $B^c=2 B.$ As a
consequence, we obtain that any $J$-invariant Gaussian measure is
uniquely determined by its complex covariation operator.

{\bf Remark.} (The origin of complex numbers) In our approach the
complex structure of QM has a natural physical explanation. The
prequantum classical field $\psi(x)$ (``background field'') is a
vector field, so $\psi(x)$ has two real components $q(x)$ and
$p(x).$ And these components are coupled in such a way that physical
variables of the $\psi$-field, $f=f(q,p),$ are $J$-invariant. Second
derivatives of such functionals are $J$-invariant ${\bf R}$-linear
symmetric operators, $f^{\prime\prime}(0)\in {\cal L}_{\rm{symp},
s}(\Omega).$ As pointed out, this space of operators can be
represented as the space of ${\bf C}$-linear operators ${\cal
L}_s(\Omega_c).$ But QM takes into account only second derivatives
of functionals of the vector prequantum field.

As in the real case [8], we  can prove that for any operator $ A\in
{\cal L}_{\rm{symp}, s}(\Omega):$
\begin{equation}
\label{CI2} \int_\Omega <A\psi,\psi> d \rho (\psi) = \rm{Tr}
\;\rm{cov}^c \rho \;A.
\end{equation}
 We pay attention that the trace is considered with respect to the complex
inner product. We consider now the one parameter family of
classical statistical models:
\begin{equation}
\label{MH} M^\alpha= ( S_{G, \rm{symp}}^\alpha(\Omega),{\cal
V}_{\rm{symp}}(\Omega)), \; \alpha\geq 0,
\end{equation}

{\bf Lemma 1.} {\it Let $f \in {\cal V}_{\rm{symp}}(\Omega)$ and let
$\rho \in S_{G, \rm{symp}}^\alpha(\Omega).$ Then the following
asymptotic equality holds:
\begin{equation}
\label{ANN3} <f>_\rho =  \frac{\alpha}{2} \; \rm{Tr}\; D^c \;
f^{\prime \prime}(0) + o(\alpha), \; \alpha \to 0,
\end{equation}
where the operator $D^c= \rm{cov}^c \; \rho/\alpha.$ Here
\begin{equation}
\label{OL} o(\alpha) = \alpha^2 R(\alpha, f, \rho),
\end{equation}
where $\vert R(\alpha,f,\rho)\vert \leq c_f\int_\Omega  e^{r_f \Vert
\psi \Vert}d\rho_{D^c} (\psi).$ }

\medskip
Here $\rho_{D^c}$ is the Gaussian measure with zero mean value and
the complex covariation operator $D^c.$

We see that the classical average (computed in the model $M^\alpha=
( S_{G, \rm{symp}}^\alpha(\Omega),{\cal V}_{\rm{symp}}(\Omega))$ by
using the measure-theoretic approach) is coupled through
(\ref{ANN3}) to the quantum average (computed in the model
$N_{\rm{quant}} =({\cal D}(\Omega_c),$ ${\cal L}_{{\rm
s}}(\Omega_c))$ by the von Neumann trace-formula).

The equality (\ref{ANN3}) can be used as the motivation for defining
the following classical $\to$ quantum map $T$ from the classical
statistical model $M^\alpha= ( S_{G, \rm{symp}}^\alpha,{\cal
V}_{\rm{symp}})$ onto the quantum statistical model
$N_{\rm{quant}}=({\cal D}, {\cal L}_{{\rm s}}):$
\begin{equation}
\label{Q20} T: S_{G, \rm{symp}}^\alpha(\Omega) \to {\cal
D}(\Omega_c), \; \; D^c=T(\rho)=\frac{\rm{cov}^c \; \rho}{\alpha}
\end{equation}
(the Gaussian measure $\rho$ is represented by the density matrix
$D^c$ which is equal to the complex covariation operator of this
measure normalized by  $\alpha$);
\begin{equation}
\label{Q30}T: {\cal V}_{\rm{symp}}(\Omega) \to {\cal L}_{{\rm
s}}(\Omega_c), \; \; A_{\rm quant}= T(f)= \frac{1}{2}
f^{\prime\prime}(0).
\end{equation}
Our previous considerations can be presented as

\medskip

{\bf Theorem 1.} {\it The one parametric family of classical
statistical models $M^\alpha= ( S_{G,
\rm{symp}}^\alpha(\Omega),{\cal V}_{\rm{symp}}(\Omega))$ provides
dequantization of the quantum model $N_{\rm{quant}} =({\cal
D}(\Omega_c),$ ${\cal L}_{{\rm s}}(\Omega_c))$ through the pair of
maps (\ref{Q20}) and (\ref{Q30}). The classical and quantum averages
are coupled by the asymptotic equality (\ref{ANN3}).}

\section{PURE STATES}

Let  $\Psi=u + iv \in \Omega_c, $ so $u \in Q, v \in P$ and let
$||\Psi||=1.$ By using the conventional terminology of quantum
mechanics we say that such a normalized vector of the complex
Hilbert space $\Psi$ represents a {\it pure quantum state.} By
Born's interpretation of the wave function a pure state $\Psi$
determines the statistical state with the density matrix:
\begin{equation}
\label{DM}D_\Psi=\Psi \otimes \Psi
\end{equation}
This Born's interpretation of the $\Psi$ -- which is, on one hand,
the pure state  (normalized vector $\Psi \in \Omega_c)$ and, on the
other hand, the statistical state $D_\Psi$ -- was the root of
appearance in QM such a notion as individual (or irreducible)
randomness. Such a randomness could not be reduced to classical
ensemble randomness, see von Neumann [11].

In our approach the density matrix $D_\Psi$ has nothing to do with
the individual state (classical field). The density matrix $D_\Psi$
is the image of the classical statistical state -- the $J$-invariant
Gaussian measure $\rho_\Psi\equiv \rho_{B_\Psi}$ on the phase space
that has the zero mean value and the complex covariation operator
$$
B_\Psi= \alpha D_\psi.
$$

\medskip

{\bf PCSFT-interpretation of pure states.} {\it There are no ``pure
quantum states.'' States that are interpreted in the conventional
quantum formalism as pure states, in fact, represent $J$-invariant
Gaussian measures having two dimensional supports. Such states can
be imagined as fluctuations of fields concentrated on two
dimensional real planes of the infinite dimensional state
phase-space.}

\section{SCHR\"ODINGER'S DYNAMICS}

States of systems with the infinite number of degrees of freedom -
classical fields -- are represented by points $\psi=(q, p) \in \Om;$
evolution of a state is described by the Hamiltonian equations. We
consider a quadratic Hamilton function: ${\cal H}(q, p)=\frac{1}{2}
({\bf H} \psi,\psi),$ where ${\bf H}: \Om \to \Om$ is an arbitrary
symmetric (bounded) operator; the Hamiltonian equations have the
form: $\dot q= {\bf H}_{21}q + {\bf H}_{22} p,  \; \;   \dot p=-(
{\bf H}_{11}q +{\bf H}_{12}p),$ or
\begin{equation}
\label{Y} \dot \psi= \left( \begin{array}{ll}
\dot q\\
\dot p
\end{array}
\right )=J{\bf H} \psi
\end{equation}
(Thus quadratic Hamilton functions induce linear Hamilton
equations.) From (\ref{Y}) we get $\psi(t)= U_t \psi, \; \;
\mbox{where} \; U_t=e^{J {\bf H} t}.$ The map $U_t\psi$ is a linear
Hamiltonian flow on the phase space $\Omega.$ Let us consider an
operator ${\bf H} \in {\cal L}_{\rm symp, s} (\Omega)$: ${\bf H}=
\left(
\begin{array}{ll}
R&T\\
-T&R
\end{array}
\right).$ This operator defines the quadratic  Hamilton function $
{\cal H}(q, p)=\frac{1}{2}[(R p, p) + 2 (Tp, q) + (Rq, q)], $ where
$R^*=R , \; \; T^*=-T.$ Corresponding Hamiltonian equations have the
form $$\dot q=Rp-Tq, \;  \dot p=-(Rq + Tp).$$ We pay attention that
for a  $J$-invariant Hamilton function, the Hamiltonian flow $U_t
\in {\cal L}_{\rm{symp}}(\Omega).$ By considering the complex
structure on the infinite-dimensional phase space $\Omega$ we write
the Hamiltonian equations (\ref{Y}) in the form of the Sch\"odinger
equation on $\Omega_c:$
$$i  \frac{d \psi}{d t} = {\bf H} \psi;$$
its solution has the following complex representation: $\psi(t)=U_t
\psi, \; \; U_t=e^{-i{\bf H} t}.$ We consider the Planck system of
units in that $h=1.$ This is {\it the complex representation of
flows corresponding to quadratic $J$-invariant Hamilton functions.}

By choosing $H=L_2( {\bf R}^n)$ we see that the interpretation of
the solution of this equation coincides with the original
interpretation of Schr\"odinger -- this is a classical field
$\psi(t,x)=(q(t,x), p(t,x).$

{\bf Example 1.} Let us consider an important class of Hamilton
functions
\begin{equation}
\label{HF} {\cal H} (q, p)=\frac{1}{2}[(Rp, p)+(Rq, q)],
\end{equation}
where $R$ is a symmetric operator. The corresponding Hamiltonian
equations have the form:
\begin{equation}
\label{HF1} \dot q=Rp, \; \dot p=-Rq.
\end{equation}
We now choose $H=L_2(\br^3),$ so $q(x)$ and $p(x)$ are components of
the vector-field $\psi(x)=(q(x), p(x)).$ We can call fields $q(x)$
and $p(x)$ {\it mutually inducing.} The field $p(x)$ induces
dynamics of the field $q(x)$ and vice versa, cf. with electric and
magnetic components, $q(x)=E(x)$ and $p(x)=B(x),$ of the
electromagnetic field, cf. Einstein and Infeld [15], p. 148: {\small
``Every change of an electric field produces a magnetic field; every
change of this magnetic field produces an electric field;  every
change of ..., and so on.''} We can write the form (\ref{HF}) as
%\begin{equation}
%\label{HF2}
${\cal H} (q, p)=\frac{1}{2} \int_{\br^6} R(x, y) [q (x) q(y) + p(x)
p(y)] dx dy$
%\end{equation}
or
%\begin{equation}
%\label{HF3}
${\cal H}(\psi)=\frac{1}{2} \int_{\br^6} R(x, y) \psi (x) \bar{\psi}
(y) dx dy ,$
%\end{equation}
where $R (x, y)=R(y, x)$ is in general a distribution on $\br^6.$ We
call such a  kernel $R(x, y)$  a {\it self-interaction potential}
for the background field $\psi(x)=(q(x), p(x)).$ We pay attention
that $R(x, y)$ induces a self-interaction of each component of the
$\psi(x),$ but there is no cross-interaction between components
$q(x)$ and $p(x)$ of the vector-field $\psi(x).$

\section{STATIONARY PURE STATES AS REPRESENTATION OF INVARIANT
GAUSSIAN MEASURES FOR SCHR\"ODINGER'S DYNAMICS}

All Gaussian measures considered in this section are supposed to be
$J$-invariant.

As we have seen in section?, so called pure states $\Psi,
||\Psi||=1,$ are just labels for Gaussian measures concentrated on
one dimensional (complex) subspaces $\Omega_\Psi$ of the
infinite-dimensional phase-space $\Omega.$ In this section we study
the case of so called {\it stationary (pure) states} in more detail.
The $\alpha$-scaling does not play any role in present
considerations. Therefore we shall not take it into account. We
consider a pure state $\Psi, ||\Psi||=1,$ as the label for the
Gaussian measure  $\nu_\Psi$ having the zero mean value and the
covariation operator ${\rm {cov}}^c \nu_\Psi=\Psi \otimes \Psi.$

\medskip

{\bf Theorem 2.} {\it Let $\nu$ be a Gaussian measure (with zero
mean value) concentrated on the one-dimensional (complex) space
corresponding to a normalized vector $\Psi$. Then $\nu$ is invariant
with respect to the unitary dynamics $U_t=e^{-it {\bf H}},$ where
${\bf H}: \Omega \to \Omega$ is a bounded self-adjoint operator, iff
$\Psi$ is an eigenvector of $\bf H.$}

{\bf Proof.} A). Let ${\bf H} \Psi= \lambda \Psi.$ The Gaussian
measure  $U_t^* \nu$ has the covariation operator $B_t^c= U_t (\Psi
\otimes \Psi) U_t^*= U_t \Psi \otimes U_t \Psi= e^{-it\lambda} \Psi
\otimes e^{-it \lambda} \Psi= \Psi \otimes \Psi.$ Since all measures
under consideration are Gaussian, this implies that  $U_t^* \nu=
\nu.$ Thus $\nu$ is an invariant measure.

B). Let $U_t^* \nu=\nu$ and $\nu=\nu_\Psi$ for some $\Psi,
||\Psi||=1.$ We have that $U_t \Psi \otimes U_t\Psi=\Psi \otimes
\Psi.$ Thus, for any $\psi_1, \psi_2 \in \Omega,$ we have
$$
<\psi_1,U_t \Psi><U_t \Psi, \psi_2>=<\psi_1, \Psi><\Psi, \psi_2>.
$$
Let us set $\psi_2=\Psi.$ We obtain: $<\psi_1, \overline{c(t)} U_t
\Psi>=<\psi_1, \Psi>,$ where $c(t)=<U_t \Psi, \Psi>. $ Thus
$\overline{c(t)} U_t \Psi=\Psi.$ We pay attention that
$c(0)=||\Psi||^2=1.$ Thus $ \overline{c^\prime (0)} \Psi -i {\bf H}
\Psi=0,$ or ${\bf H} \Psi=-i  \overline{c^\prime (0)} \Psi.$ Thus
$\Psi$ is an eigenvector of $\bf H$ with the eigenvalue $-i
\overline{c^\prime (0)}.$ We remark that $c^\prime (0)=-i <{\bf H}
\Psi, \Psi>; $ so $ \overline{c^\prime (0)}=i <{\bf H} \Psi, \Psi>.$
Hence, $\lambda=-i  \overline{c^\prime (0)}=<{\bf H}, \Psi, \Psi>.$

\medskip

{\bf Conclusion.} {\it {Stationary states of quantum Hamiltonian
represented by a bounded self-adjoint operator $\bf H$  are just
labels for Gaussian one-dimensional measures (with the zero mean
value) that are invariant with respect to the Schr\"odinger dynamics
$U_t=e^{-it {\bf H}}$.}}

\medskip

We now describe all possible Gaussian measures which are
$U_t$-invariant.

{\bf Theorem 3.} {\it {Let $\bf H$ be a bounded self-adjoint
operator with purely discrete nondegenerate spectrum: ${\bf H}
\Psi_k=\lambda_k \Psi_k,$ so $\{\Psi_k\}$ is an orthonormal basis
consisting of eigenvectors of $\bf H.$ Then any $U_t$-invariant
Gaussian measure $\nu$ (with the zero mean value) has the covariance
operator of the form:

\begin{equation}
\label{COV} B^c=\sum_{k=1}^\infty c_k \Psi_k \otimes \Psi_k, c_k\geq
0,
\end{equation}

and vice versa.}}

{\bf Proof.} A). Let ${\rm {cov}}^c \nu=B^c$ has the form
(\ref{COV}). Then
$$
{\rm {cov}}^c U_t^* \nu= U_t B U_t^*= \sum_{k=1}^\infty c_k e^{-i
\lambda_k t} \Psi_k \otimes e^{-i\lambda_k t} \Psi_k={\rm {cov}}^c
\nu=B^c.
$$
Since measures are Gaussian, this implies that $U_t^* \nu=\nu$ for
any $t.$

B). Let $U_t^* \nu=\nu$ for any $t.$ We remark that any covariation
operator $B^c$ can be represented in the form:
$$
B^c=\sum_{k=1}^\infty <B \Psi_k, \Psi_k> \Psi_k \otimes \Psi_k +
\sum_{k \ne j} <B \Psi_k, \Psi_j> \Psi_k \otimes \Psi_j.
$$
We shall show that $<B \Psi_k, \Psi_j>=0$ for $k \ne j.$ Denote the
operator corresponding to $\sum_{k \ne j}$ by Z. We have
$$
<U_t Z U_t \psi_1, \psi_2>= \sum_{k \ne j} <B \Psi_k, \Psi_j>
e^{it(\lambda_j - \lambda_k)} <\Psi_k, \psi_2> <\psi_1, \Psi_j>= <Z
\psi_1, \psi_2>.$$ Set $\psi_1=\Psi_j, \psi_2= \Psi_k.$ Then
$$<U_t Z U_t^* \Psi_j, \Psi_k>= <B \Psi_k, \Psi_j> e^{it(\lambda_j -
\lambda_k)}=<B \Psi_k, \Psi_j>.
$$
Thus $<B \Psi_k, \Psi_j>=0, k \ne
j.$

\section{STABILITY OF HYDROGEN ATOM IN PCSFT}

As we have seen, in PCSFT so called stationary (pure) states of
quantum mechanics are just labels for Gaussian measures (which are
$J$-invariant and have zero mean value) that are $U_t$-invariant. We
now apply our standard $\alpha$-scaling argument and we see that a
stationary state $\Psi$ is a label for the Gaussian measure
$\rho_\Psi$ with ${\rm cov}^c \rho_\Psi=\alpha \Psi \otimes \Psi.$
This measure is concentrated on one-dimensional (complex) subspace
$\Omega_\Psi$ of phase space $\Omega.$ Therefore each realization of
an element of the Gaussian ensemble of classical fields
corresponding to the statistical state $\rho_\Psi$ gives us the
field of the shape $\Psi(x),$ but magnitudes of these fields vary
from one realization to another. But by Chebyshov inequality
probability that ${\cal E}(\Psi)=\int_{\br^3}|\Psi(x)|^2 dx$ is
large is negligibly small.

Thus we have Gaussian fluctuations of very small magnitudes of the
same shape $\Psi(x).$ In PCSFT a stationary quantum state can not be
identified with a stationary classical field, but only with an
ensemble of fields having the same shape $\Psi(x).$ Let us now
compare descriptions of dynamics of electron in hydrogen atom given
by quantum mechanics and our prequantum field theory.

In quantum mechanics stationary bound states of hydrogen atom are of
the form: $$ \Psi_{nlm}(r, \theta, \phi)=c_{n,l} R^l L_{n + l}^{2l +
1} (R) e^{-R/2} Y_l^m (\theta, \phi),$$ where $R=\frac{2r}{n a_0},$
and $a_0=\frac{h^2}{\mu e^2}$ is a characteristic length for the
atom (Bohr radius). We are mainly interested in the presence of the
component $e^{-R/2}.$

In PCSFT this stationary bound state is nothing else, but the label
for the Gaussian measure $\rho_{\Psi_{nlm}}$ which is concentrated
on the subspace $\Omega_{\Psi_{nlm}}.$ Thus PCSFT says that
``electron in atom" is nothing else as Gaussian fluctuations of the
classical field $\Psi_{nlm} (r, \theta, \phi):$
\begin{equation}
\label{FL} \psi_{nlm} (r, \theta, \phi; \omega)=\gamma (\omega)
\Psi_{nlm} (r, \theta, \phi),
\end{equation}
where $\gamma(\omega)$ is the C-valued Gaussian random variable:
$E\gamma=0, E|\gamma|^2=\alpha.$

The intensiveness of the field $\Psi_{nlm} (r, \theta, \phi,
\omega)$ varies, but the shape is the same. Therefore this random
field does not produce any significant effect for large $R$ (since
$e^{-R/2}$ eliminates such effects).

Thus in PCSFT the hydrogen atom stable, since the prequantum random
fields $\psi_{nlm} (r, \theta, \phi; \omega)$ have a special shape
(decreasing exponentially $R \to \infty).$

This is a good place to discuss the role of physical space
represented by ${\bf R}^3$ in our model. In PCSFT the real physical
space is Hilbert space. If we choose the realization $H=L_2({\bf
R}^3),$ then we obtain the realization of $H$ as the space of
classical fields on ${\bf R}^3.$  So conventional space ${\bf R}^3$
appears only through this special representation of Hilbert
configuration space. Dynamics in ${\bf R}^3$ in just a shadow of
dynamics in the space of fields. However, we can choose other
representations of Hilbert configuration space. In this way we shall
obtain classical fields defined on other ``physical spaces.''

\section{APPENDIX}

{\bf Proof of Lemma 1.} In the Gaussian integral $\int_\Omega f(
\psi) d\rho( \psi)$ we make the scaling:
\begin{equation}
\label{LHT}  \psi \to \frac{ \psi}{\sqrt{\alpha}} .
\end{equation}
 We denote the image of the
measure $\rho$ under this change of variables by $\rho_{D^c},$ since
the latter measure (which is also Gaussian) has the complex
covariation operator $D^c.$ We have:
\begin{equation}
\label{ANN1} <f>_{\rho}= \int_\Omega f(\sqrt{\alpha}  \psi)
d\rho_{D^c} ( \psi)= \frac{\alpha}{2} \int_\Omega (f^{\prime
\prime}(0)\psi, \psi) \; d\rho_{D^c}(\psi) + \alpha^2
R(\alpha,f,\rho{D^c}),
\end{equation}
where
$$
R(\alpha,f,\rho)= \int_\Omega g(\alpha,f; \psi) d\rho_{D^c} (\psi),
g(\alpha,f; \psi)= \sum_{n=4}^\infty \frac{\alpha^{n/2-2}}{n!}
 f^{(n)}(0)( \psi, ..., \psi).
$$
 We pay attention that
$$ \int_\Omega (f^\prime(0),  \psi) d\rho_{D^c}(
\psi)=0,\; \;\; \; \int_\Omega
f^{\prime\prime\prime}(0)(\psi,\psi,\psi) d\rho_{D^c}( \psi)=0,
$$
because the mean value of $\rho$  (and, hence, of $\rho_{D^c})$ is
equal to zero. Since $\rho\in S_{G, \rm{symp}}^\alpha(\Omega),$ we
have $\rm{Tr} \; {D^c} = 1.$ We now estimate the rest term
$R(\alpha,f,\rho).$ We recall the following inequality for functions
of the exponential growth:
\begin{equation}
\label{ZO1} \Vert f^{(n)} (0) \Vert \leq  c\;  r^n, \; n=0,1, 2,...
\end{equation}
This inequality is well known for analytic functions of the
exponential growth $f: {\bf C}^n \to {\bf C}.$ It was generalized to
infinite-dimensional case in [16].

By using this inequality we have for $\alpha \leq 1:$
$$
\vert g(\alpha,f; \psi)\vert  = \sum_{n=4}^\infty  \frac{\Vert
f^{(n)} (0) \Vert \Vert \psi \Vert^n}{n!} \leq c_f \sum_{n=4}^\infty
\frac{r_f^n  \Vert \psi \Vert^n}{n!}= C_f e^{r_f \Vert \psi \Vert}.
$$
Thus: $ \vert R(\alpha,f,\rho)\vert \leq c_f\int_\Omega  e^{r_f
\Vert \psi \Vert}d\rho_{D^c} (\psi). $ We obtain:
\begin{equation}
\label{ANN2} <f>_\rho=  \frac{\alpha}{2} \int_\Omega (f^{\prime
\prime}(0)\psi, \psi) \; d\rho_{D^c}(\psi) + o(\alpha), \; \alpha
\to 0.
\end{equation}
By using the equalities (\ref{CI1}) and (\ref{CI2}) we finally come
the asymptotic equality (\ref{ANN3}).

\medskip

{\bf REFERENCES}

\medskip

1. L. Marchildon, ``The epistemic view of quantum states and the
ether'', quant-ph/0510120; {\it Found. Phys.} {\bf 34}, 1453 (2004);
{\it Quantum mechanics: from basic principles to numerical methods
and applications} (Springer, Berlin-Heidelberg-New York, 2002)

2. G. `t Hooft, ``Quantum Mechanics and Determinism,''
hep-th/0105105; G. `t Hooft,``Determinism beneath Quantum
Mechanics,'' quant-ph/0212095.

3. W. M. De Muynck, {\it Foundations of quantum mechanics, an
empiricists approach} (Kluwer, Dordrecht, 2002); ``Interpretations
of quantum mechanics, and interpretations of violations of Bell's
inequality'', in {\it Foundations of Probability and Physics}, A.
Yu. Khrennikov, ed, {\it Q. Prob. White Noise Anal.} {\bf  13},  95
(2001), pp. 95-114.

4.  A. Plotnitsky,  {\it Found. Phys.} {\bf 33}, 1649 (2003); {\it
The knowable and unknowable (Modern science, nonclassical thought,
and the ``two cultures)} (Univ. Michigan Press, 2002); ``Quantum
atomicity and quantum information: Bohr, Heisenberg, and quantum
mechanics as an information theory'', in {\it Quantum theory:
reconsideration of foundations},  A. Yu. Khrennikov,ed.( V\"axj\"o
Univ. Press,  2002), pp. 309-343.

5.  A. Yu. Khrennikov (editor), {\it Foundations of Probability and
Physics,} Q. Prob. White Noise Anal.,  13,  WSP, Singapore, 2001;
{\it Quantum Theory: Reconsideration of Foundations,} Ser. Math.
Modeling, 2, V\"axj\"o Univ. Press,  2002;  {\it Foundations of
Probability and Physics}-2, Ser. Math. Modeling, 5, V\"axj\"o Univ.
Press,  2003; {\it Quantum Theory: Reconsideration of
Foundations}-2,  Ser. Math. Modeling, 10, V\"axj\"o Univ. Press,
2004; Proceedings of Conference {\it Foundations of Probability and
Physics-3,} American Institute of Physics, Ser. Conference
Proceedings, {\bf 750}, 2005.

6. K. Hess and W. Philipp, {\it Proc. Nat. Acad. Sc.} {\bf 98},
14224 (2001); {\it Europhys. Lett.} {\bf 57}, 775 (2002); ``Bell's
theorem: critique of proofs with and without inequalities'', in {\it
Foundations of Probability and Physics}-3, A. Yu. Khrennikov, ed.,
AIP Conference Proceedings, 2005, pp. 150-157.

7. A. Yu. Khrennikov,  {\it Annalen  der Physik} {\bf 12}, 575
(2003); {\it J. Phys.A: Math. Gen.} {\bf 34}, 9965 (2001); {\it Il
Nuovo Cimento} B {\bf 117},   267 (2002); {\it J. Math. Phys.} {\bf
43} 789 (2002); Ibid {\bf 45}, 902 (2004); {\it Doklady Mathematics}
{\bf 71},  363 (2005).

8. A. Yu. Khrennikov,  ``Prequantum classical statistical model with
infinite dimensional phase-space'', {\it J. Phys. A: Math. Gen.},
{\bf 38}, 9051-9073 (2005);   Generalizations of Quantum Mechanics
Induced by Classical Statistical Field Theory. {\it Found. Phys.
Letters}, {\bf 18}, 637-650.

9. E. Nelson , {\it Quantum fluctuation} (Princeton Univ. Press,
Princeton, 1985); L. de la Pena and A. M. Cetto, {\it The Quantum
Dice: An Introduction to Stochastic Electrodynamics} (Kluwer,
Dordrecht, 1996); T. H. Boyer, {\it A Brief Survey of Stochastic
Electrodynamics} in Foundations of Radiation Theory and Quantum
Electrodynamics,  A. O. Barut, ed. (Plenum, New York, 1980); T. H.
Boyer, Timothy H., {\it Scientific American}, pp. 70-78, Aug 1985;
L. De La Pena, {\it Found. Phys.} {\bf 12}, 1017 (1982); {\it J.
Math. Phys.} {\bf 10}, 1620 (1969); L. De La Pena, A. M. Cetto, {\it
Phys. Rev. D} {\bf 3}, 795 (1971).

10. P. A. M.  Dirac, {\it The Principles of Quantum Mechanics}
(Oxford Univ. Press, 1930).

11. J. von Neumann, {\it Mathematical foundations of quantum
mechanics} (Princeton Univ. Press, Princeton, N.J., 1955).

12. E. Binz, R. Honegger, A. Rieckers,  ``Field-theoretic Weyl
Quantization as a Strict and Continuous Deformation Quantization,''
{\it Ann. Henri Poincaré} {\bf 5}, 327 (2004).

13. L. Arnold,  {\it Random dynamical systems} (Springer Verlag,
Berlin-New York-Heidelberg, 1998).

14. J. S. Bell, {\it Speakable and unspeakable in quantum mechanics}
(Cambridge Univ. Press, 1987).

15. A. Einstein and L. Infeld, {\it The evolution of Physics. From
early concepts to relativity and quanta} (Free Press, London, 1967).

16. A. Yu. Khrennikov, Equations with infinite-dimensional
pseudo-differential operators. Dissertation for the degree of
candidate of phys-math. sc., Dept. Mechanics-Mathematics, Moscow
State University, Moscow, 1983.

\end{document}